\documentclass[ALICE,manyauthors]{cernphprep}
\usepackage[comma,square,numbers,sort&compress]{natbib}
\usepackage{hyperref}
\usepackage{lineno}
\usepackage{xspace}
\usepackage{xcolor}
\usepackage{orcidlink}
\begin{document}
%%%%%%%%%%%%%%%%%%%%%%%%%%%%%%%%%%%%%%%%%%%%%%%%%%
% These are some new commands that may be useful 
% for paper writing in general. If other newcommands
% are needed for your specific paper, please feel 
% free to add here. 
%
% The currently available commands are organized in: 
% 1) Systems
% 2) Quantities
% 3) Energies and units
% 4) Detectors
% 5) particle species 
%%%%%%%%%%%%%%%%%%%%%%%%%%%%%%%%%%%%%%%%%%%%%%%%%%

% 1) SYSTEMS 
\newcommand{\pp}           {pp\xspace}
\newcommand{\ppbar}        {\mbox{$\mathrm {p\overline{p}}$}\xspace}
\newcommand{\XeXe}         {\mbox{Xe--Xe}\xspace}
\newcommand{\PbPb}         {\mbox{Pb--Pb}\xspace}
\newcommand{\pA}           {\mbox{pA}\xspace}
\newcommand{\pPb}          {\mbox{p--Pb}\xspace}
\newcommand{\AuAu}         {\mbox{Au--Au}\xspace}
\newcommand{\dAu}          {\mbox{d--Au}\xspace}

% 2) QUANTITIES 
\newcommand{\s}            {\ensuremath{\sqrt{s}}\xspace}
\newcommand{\snn}          {\ensuremath{\sqrt{s_{\mathrm{NN}}}}\xspace}
\newcommand{\pt}           {\ensuremath{p_{\rm T}}\xspace}
\newcommand{\jetpt}        {\ensuremath{p_{\rm T,ch\:jet}}\xspace}
\newcommand{\meanpt}       {$\langle p_{\mathrm{T}}\rangle$\xspace}
\newcommand{\ycms}         {\ensuremath{y_{\rm CMS}}\xspace}
\newcommand{\ylab}         {\ensuremath{y_{\rm lab}}\xspace}
\newcommand{\etarange}[1]  {\mbox{$\left | \eta \right |~<~#1$}}
\newcommand{\yrange}[1]    {\mbox{$\left | y \right |~<~#1$}}
\newcommand{\dndy}         {\ensuremath{\mathrm{d}N_\mathrm{ch}/\mathrm{d}y}\xspace}
\newcommand{\dndeta}       {\ensuremath{\mathrm{d}N_\mathrm{ch}/\mathrm{d}\eta}\xspace}
\newcommand{\avdndeta}     {\ensuremath{\langle\dndeta\rangle}\xspace}
\newcommand{\dNdy}         {\ensuremath{\mathrm{d}N_\mathrm{ch}/\mathrm{d}y}\xspace}
\newcommand{\Npart}        {\ensuremath{N_\mathrm{part}}\xspace}
\newcommand{\Ncoll}        {\ensuremath{N_\mathrm{coll}}\xspace}
\newcommand{\dEdx}         {\ensuremath{\textrm{d}E/\textrm{d}x}\xspace}
\newcommand{\RpPb}         {\ensuremath{R_{\rm pPb}}\xspace}

% 3) ENERGIES, UNITS
\newcommand{\nineH}        {$\sqrt{s}~=~0.9$~Te\kern-.1emV\xspace}
\newcommand{\seven}        {$\sqrt{s}~=~7$~Te\kern-.1emV\xspace}
\newcommand{\twoH}         {$\sqrt{s}~=~0.2$~Te\kern-.1emV\xspace}
\newcommand{\twosevensix}  {$\sqrt{s}~=~2.76$~Te\kern-.1emV\xspace}
\newcommand{\five}         {$\sqrt{s}~=~5.02$~Te\kern-.1emV\xspace}
\newcommand{\twosevensixnn}{$\sqrt{s_{\mathrm{NN}}}~=~2.76$~Te\kern-.1emV\xspace}
\newcommand{\fivenn}       {$\sqrt{s_{\mathrm{NN}}}~=~5.02$~Te\kern-.1emV\xspace}
\newcommand{\Bfivenn}       {$\mathbf{\sqrt{s_{\mathrm{NN}}}~=~5.02}$~Te\kern-.1emV\xspace}
\newcommand{\LT}           {L{\'e}vy-Tsallis\xspace}
\newcommand{\GeVc}         {Ge\kern-.1emV/$c$\xspace}
\newcommand{\MeVc}         {Me\kern-.1emV/$c$\xspace}
\newcommand{\TeV}          {Te\kern-.1emV\xspace}
\newcommand{\GeV}          {Ge\kern-.1emV\xspace}
\newcommand{\MeV}          {Me\kern-.1emV\xspace}
\newcommand{\GeVmass}      {Ge\kern-.2emV/$c^2$\xspace}
\newcommand{\MeVmass}      {Me\kern-.2emV/$c^2$\xspace}
\newcommand{\lumi}         {\ensuremath{\mathcal{L}}\xspace}

% 4) DETECTORS 
\newcommand{\ITS}          {\rm{ITS}\xspace}
\newcommand{\TOF}          {\rm{TOF}\xspace}
\newcommand{\ZDC}          {\rm{ZDC}\xspace}
\newcommand{\ZDCs}         {\rm{ZDCs}\xspace}
\newcommand{\ZNA}          {\rm{ZNA}\xspace}
\newcommand{\ZNC}          {\rm{ZNC}\xspace}
\newcommand{\SPD}          {\rm{SPD}\xspace}
\newcommand{\SDD}          {\rm{SDD}\xspace}
\newcommand{\SSD}          {\rm{SSD}\xspace}
\newcommand{\TPC}          {\rm{TPC}\xspace}
\newcommand{\TRD}          {\rm{TRD}\xspace}
\newcommand{\VZERO}        {\rm{V0}\xspace}
\newcommand{\VZEROA}       {\rm{V0A}\xspace}
\newcommand{\VZEROC}       {\rm{V0C}\xspace}
\newcommand{\Vdecay} 	   {\ensuremath{V^{0}}\xspace}

% 4) PARTICLE SPECIES 
\newcommand{\ee}           {\ensuremath{e^{+}e^{-}}} 
\newcommand{\pip}          {\ensuremath{\pi^{+}}\xspace}
\newcommand{\pim}          {\ensuremath{\pi^{-}}\xspace}
\newcommand{\kap}          {\ensuremath{\rm{K}^{+}}\xspace}
\newcommand{\kam}          {\ensuremath{\rm{K}^{-}}\xspace}
\newcommand{\pbar}         {\ensuremath{\rm\overline{p}}\xspace}
\newcommand{\kzero}        {\ensuremath{{\rm K}^{0}_{\rm{S}}}\xspace}
\newcommand{\lmb}          {\ensuremath{\Lambda}\xspace}
\newcommand{\almb}         {\ensuremath{\overline{\Lambda}}\xspace}
\newcommand{\Om}           {\ensuremath{\Omega^-}\xspace}
\newcommand{\Mo}           {\ensuremath{\overline{\Omega}^+}\xspace}
\newcommand{\X}            {\ensuremath{\Xi^-}\xspace}
\newcommand{\Ix}           {\ensuremath{\overline{\Xi}^+}\xspace}
\newcommand{\Xis}          {\ensuremath{\Xi^{\pm}}\xspace}
\newcommand{\Oms}          {\ensuremath{\Omega^{\pm}}\xspace}
\newcommand{\degree}       {\ensuremath{^{\rm o}}\xspace}

%%%%%%%%%%%%%%%  Title page %%%%%%%%%%%%%%%%%%%%%%%%
\begin{titlepage}
% the dates below correspond to CERN approval
% please don't touch: EB chairs will take care
\PHyear{2023}       % required, will be obtained from CERN
\PHnumber{154}      % required, will be obtained from CERN
\PHdate{26 July}  % required, will be obtained from CERN
%%%%%%%%%%%%%%%%%%%%%%%%%%%%%%%%%%%%%%%%%%%%%%%%%%%%

%%% Put your own title + short title here:
\title{Modification of charged-particle jets in event-shape engineered \PbPb collisions at \Bfivenn }
\ShortTitle{Jets and Event-Shape Engineering}   % appears on left page headers

%%% Do not change the next lines
\Collaboration{ALICE Collaboration\thanks{See Appendix~\ref{app:collab} for the list of collaboration members}}
\ShortAuthor{ALICE Collaboration} % appears on right page headers, do not change

\begin{abstract}
Charged-particle jet yields have been measured in semicentral \PbPb collisions at center-of-mass energy per nucleon--nucleon collision \fivenn with the ALICE detector at the LHC. These yields are reported as a function of the jet transverse momentum, and further classified by their angle with respect to the event plane and the event shape, characterized by ellipticity, in an effort to study the path-length dependence of jet quenching. Jets were reconstructed at midrapidity from charged-particle tracks using the anti-$k_{\rm T}$ algorithm with resolution parameters $R$ = 0.2 and 0.4, with event-plane angle and event-shape values determined using information from forward scintillating detectors. The results presented in this letter show that, in semicentral \PbPb collisions, there is no significant difference between jet yields in predominantly isotropic and elliptical events. However, out-of-plane jets are observed to be more suppressed than in-plane jets. Further, this relative suppression is greater for low transverse momentum ($<$ 50 GeV/\textit{c}) $R$ = 0.2 jets produced in elliptical events, with out-of-plane to in-plane jet-yield ratios varying up to 5.2$\sigma$ between different event-shape classes. These results agree with previous studies indicating that jets experience azimuthally anisotropic suppression when traversing the QGP medium, and can provide additional constraints on the path-length dependence of jet energy loss.

\end{abstract}

\end{titlepage}

\setcounter{page}{2} %please do not remove this line

%%%%%%%%%%%%%%%%%%%%%%%%%%%%%%%%
% begin main text
%%%%%%%%%%%%%%%%%%%%%%%%%%%%%%%%

\section{Introduction} 

At very high energy densities, ordinary hadronic matter undergoes a transition to become a strongly interacting state of deconfined quarks and gluons. This new state of matter is referred to as the quark--gluon plasma (QGP)~\cite{Wilke}. Calculations using quantum chromodynamics (QCD) on the lattice predict a crossover transition between these phases at a temperature of about $150$ MeV that can be reached in the laboratory via ultrarelativistic collisions of heavy ions~\cite{HotQCD150, Ratti:2018ksb}. Experimental studies of heavy-ion collisions thus offer a compelling opportunity to explore the properties of the strongly interacting medium, and form the main physics program of the ALICE experiment at the LHC~\cite{ALICEReview}.

Jets, sprays of hadrons resulting from high-transverse-momentum (\pt) partons produced in hard-scattering processes, are sensitive to a variety of QGP properties~\cite{Gyulassy:1990ye, JetReviewXinNian, jetreview}. Because jets are produced early in a collision, indeed much earlier than the formation of the QGP at $\tau_{\mathrm{QGP}}\sim0.5$ fm/\textit{c}, they experience its whole evolution. Jets interact with and are modified by this medium as they traverse it, resulting in a collection of effects known as jet quenching. The observation of jet quenching at both RHIC and LHC energies is therefore considered to be a main signature of QGP formation~\cite{ALICEjetRaa, STARjetRaa, CMSjetRaa, ATLASjetRaa}, and the microscopic mechanism by which jet quenching occurs has been the subject of significant theoretical and experimental investigation. Models predict that partons can lose energy collisionally and/or radiatively in the weakly-coupled limit, with radiative contributions expected to dominate in the high-\pt regime~\cite{BDMPS, GLV}. Moreover, it is predicted that there is a direct relationship between the path-length dependence of parton energy loss and the relative contributions of the different mechanisms. In a static medium, collisional and radiative energy loss are expected to have a linear and quadratic dependence on the length of plasma traversed, respectively~\cite{djordjevic, Djordjevic:2003zk}. Measuring this dependence would therefore offer a direct way to probe the underlying mechanisms of jet--medium interactions, but doing so has so far proven to be challenging. Past measurements, e.g. the dijet asymmetry~\cite{ATLASdijet, ATLASxj, CMSAj}, are heavily influenced by fluctuations in jet--medium interactions, making it difficult to extract an underlying path-length dependence~\cite{zapp}. Another such measurement, the jet $v_{\rm 2}$ (the second Fourier coefficient in the azimuthal distribution of jet momenta in the transverse plane), shows a significant azimuthal anisotropy in jet yields in \PbPb collisions~\cite{jetv2, ATLASjetv2, ATLASjetv2502}. However, medium fluctuations limit the ability to constrain the underlying physics mechanisms that drive this behavior.\par

Event-shape engineering (ESE)~\cite{ese}, a technique that classifies events according to their anisotropies using the magnitude of the reduced flow vector, offers a new experimental approach to overcome these difficulties and constrain the path-length dependence of jet energy loss~\cite{Christiansen:2016uaq}. This approach is advantageous in that it allows for the selection of events for which the thermodynamic properties are similar, but for which the spatial anisotropies vary significantly. This is done by isolating events within a centrality class that have particularly round or elliptical geometries. Previous measurements have shown that the elliptic flow coefficient $v_{2}$ of charged particles varies significantly at a fixed collision centrality~\cite{ESE276, CF_ESE, ATLAS:2015qwl}. In addition, the mean $\pt$ of the particle yields is larger for elliptical events than for isotropic events. Results using ESE in the heavy-flavor sector show similar indications~\cite{HF1, HF2}. These measurements reveal the sizeable potential that ESE has to connect observables from the soft and hard momentum scales.

Combining the precision afforded by jet measurements with the control that ESE provides to constrain the collision geometry, it is possible to learn about this interplay of physics phenomena from high to low \pt~\cite{Beattie:2022ojg}. In the analysis presented in this letter, this interplay is studied by considering an event shape in conjunction with the jet angle with respect to the event-plane $\Psi_{\rm 2}$, which is defined by the beam axis and the vector of the collision impact parameter. The distance the jet traverses through the medium when traveling parallel to the event plane (in-plane) is, on average, shorter than when it travels perpendicular to the event plane (out-of-plane). As such, the azimuthal anisotropy of the jet spectra provides initial information about the path-length dependence of parton energy loss. By then applying ESE, the relative difference between in- and out-of-plane jet path-lengths can be increased or decreased. This is especially true in the case of semicentral collisions, where the system is usually (but not necessarily) elliptical~\cite{Beattie:2022ojg}. In semicentral \PbPb collisions at $\sqrt{s_{\rm NN}} = 2.76$ TeV, the charged-particle $v_{\rm 2}$ ratio for the 30\% most elliptical events compared to the 30\% most isotropic events is $\sim$1.3~\cite{CF_ESE}. This ratio was approximated by considering the average of the charged-particle $v_{\rm 2}$ values reported in differentiated centrality and ellipticity windows. With this in consideration, comparing  in- and out-of-plane jet spectra for events with different ellipticities can reduce the contribution of medium shape fluctuations and increase understanding of the path-length dependence of jet energy loss. \par

In this letter, results of event-shape engineered jet yields in 30--50\% \PbPb collisions at \fivenn are presented. Jets were reconstructed from charged-particle tracks for resolution parameters $R=0.2$ and $R=0.4$, within a jet transverse-momentum range of $35<\jetpt<120~$GeV/\textit{c} and $40<\jetpt<120~$GeV/\textit{c}, respectively. The in-plane and out-of-plane jet yields are presented according to the ellipticity of the collision system quantified event-by-event, which allows for the exploitation of average differences in jet path length.

\section{Experimental Setup}

The ALICE experiment is a general-purpose detector located at the LHC. It is optimized to provide high momentum resolution and excellent particle identification over a broad momentum range, up to the highest multiplicities~\cite{CentralityDetermination}. The primary ALICE sub-detectors used in this analysis are the Inner Tracking System (ITS), Time Projection Chamber (TPC), and V0 detectors. For more information on the ALICE apparatus and its performance, see Refs.~\cite{ALICEExperiment, ALICE:2014sbx}.

The ITS is a silicon-based tracking detector used for reconstruction of charged tracks and primary vertex identification~\cite{ALICE:2010tia}. It consists of six layers having increasing radii around the nominal collision point. The first two layers are Silicon Pixel Detectors (SPD), followed by two layers of Silicon Drift Detectors (SDD), and two layers of Silicon Strip Detectors (SSD). The TPC is a large cylindrical gaseous detector, covering a pseudorapidity range of $|\eta|<0.9$ over the full azimuthal angle~\cite{Alme:2010ke} and providing excellent tracking performance up to high particle multiplicities and momenta. The tracks used for jet reconstruction in this analysis were measured by both the ITS and the TPC, and were accepted for $\pt > 0.15$ GeV/\textit{c} and pseudorapitidies of $|\eta|<0.9$. The tracks have a momentum resolution of $\sigma_{\pt}/\pt \sim 0.8\%$ at $\pt=1$ GeV/\textit{c}, which increases to $\sigma_{\pt}/\pt \sim 2\%$ at $\pt=10$ GeV/\textit{c}~\cite{ALICE:2014sbx}. In central Pb--Pb collisions, the tracking efficiency ranges from approximately $65\%$ to $82\%$ for increasing \pt~\cite{ALICEjetRaa}.

The V0A and V0C, scintillation detectors located at pseudorapidities  $2.8<\eta<5.1$ and $-3.7<\eta<-1.7$, respectively, were used to select the \PbPb minimum-bias and semicentral events according to their summed amplitudes~\cite{ALICE:2004ftm, ALICE:2013axi}. In this analysis, the V0C was used for calculating the reduced flow vector $q_{\rm 2}$, defined in Eq.~\ref{eq:q2}, as it is closer to midrapidity than the V0A. It can therefore produce a wider $q_{\rm 2}$ distribution, thus accessing the best separation between different event shapes. The V0A was used for calculating event-plane angles. Using these forward detectors to measure the event shape and event-plane angle minimizes the autocorrelations between these quantities and the charged-particle jets at midrapidity. Details of the $q_{\rm 2}$ and event-plane angle measurements are given in the next section.

\section{Data Analysis}

The results presented in this letter are derived from a sample of \PbPb collisions collected by the ALICE experiment during the 2018 LHC heavy-ion run. The data sample considered in this work was recorded with a semicentral trigger based on the V0 signal amplitude, which allowed for the collection of a large sample of Pb--Pb collisions in the 30--50\% centrality class~\cite{CentralityDetermination}. Only events having a primary vertex within $\pm 10~$cm of the nominal interaction point along the beam line ($z$ direction) were accepted. An additional selection criterion was applied to remove pile-up, utilizing the correlation between the number of hits in the ITS and TPC detectors. After applying these criteria, a total of approximately 54 million events were selected for this study.

Jets were reconstructed from charged-particle tracks~\cite{ALICE:2012nbx} with the FastJet anti-$k_{\rm T}$ algorithm~\cite{FastJet, antikt}. The $\pt$-scheme recombination strategy was chosen to combine tracks using their transverse momenta~\cite{recoSchemes, FastJet}. The resolution parameters $R=0.2$ and $R=0.4$ were studied, where each jet was required to contain a leading track with $5 < \pt < 100$ GeV/\textit{c}. The leading track requirement was chosen to reduce contamination from combinatorial jets. The jet axis was required to be within $|\eta_{\mathrm{jet}}|<0.9-R$, where $\eta_{\mathrm{jet}}$ is the pseudorapidity of the jet axis. Furthermore, for each jet the quantity $\Delta\varphi = \varphi_{\rm jet} - \Psi_{\rm 2}$ was calculated. This is the difference in azimuthal angle between the jet axis and the event-plane angle $\Psi_{\rm 2}$, where $\Psi_{\rm 2}$ is determined from the V0A signals. The average combinatorial background was subtracted using an area-based technique~\cite{Cacciari:2007fd, Cacciari:2010te, ALICE:2012nbx}. With this method, the background transverse-momentum density per unit area, $\rho$, was determined event-by-event after removing the two leading $k_{\rm T}$ jets~\cite{kT}. The jet energies were corrected for the underlying-event contribution by subtracting the event-averaged density multiplied by the jet area. The residual background fluctuations, together with detector effects, were then corrected on a statistical basis using a 2D Bayesian unfolding procedure~\cite{Bayesunfold, RooUnfold}. The choice to use a 2D procedure was made so as to correct for the differences in background arising from the jet angle with respect to $\Psi_{\rm 2}$, as well as to account for any correlated bin migration in $\Delta\varphi$ and \jetpt. This was done using a 4D response matrix constructed from PYTHIA 8 (Monash tune)~\cite{PYTHIA8.2, Monash} jets transported through the ALICE detector by a GEANT3-based simulation~\cite{GEANT} and embedded into real Pb–Pb events. The data was binned in \jetpt and $|\cos(\Delta\varphi)|$ for both truth- and reconstructed-level jets. Before filling the response matrix, 2\% of simulated tracks were randomly rejected before jet-finding to account for the worsened tracking efficiency in the high track-density environment of Pb--Pb collisions. This level of degradation was estimated using HIJING simulations of 0--10\% central Pb--Pb collisions~\cite{HIJING}. The 2D jet distribution was then unfolded using six iterations of the Bayesian procedure, with the PYTHIA 8 distribution used as the prior. \par

After unfolding, corrections were applied to the jet yields for the kinematic and reconstruction efficiencies. Here, the kinematic efficiency refers to the inefficiency introduced by truth-level jets that were reconstructed outside of the measured \jetpt range, thus not entering the unfolding procedure. This was computed for each bin by taking the ratio of the truth-level spectrum reconstructed in the measured range to the truth level-spectrum reconstructed within $10 < \jetpt < 200$ GeV/\textit{c}. The reconstruction efficiency accounts for truth-level jets that were not found at detector-level. Corrections were also applied to account for the event-plane resolution when the event-plane angle was considered. The reported \jetpt ranges are $35 - 120$ GeV/\textit{c} and $40 - 120$ GeV/\textit{c} for $R=0.2$ and $R=0.4$ jets, respectively. These ranges were chosen to satisfy the requirement of having a kinematic efficiency above 75\% for each generator-level \jetpt bin, as well as to ensure stability when varying the lower limit of the \jetpt range considered in the unfolding procedure.

\begin{figure}[tb]
    \begin{center}
    \includegraphics[width = 0.5\textwidth]{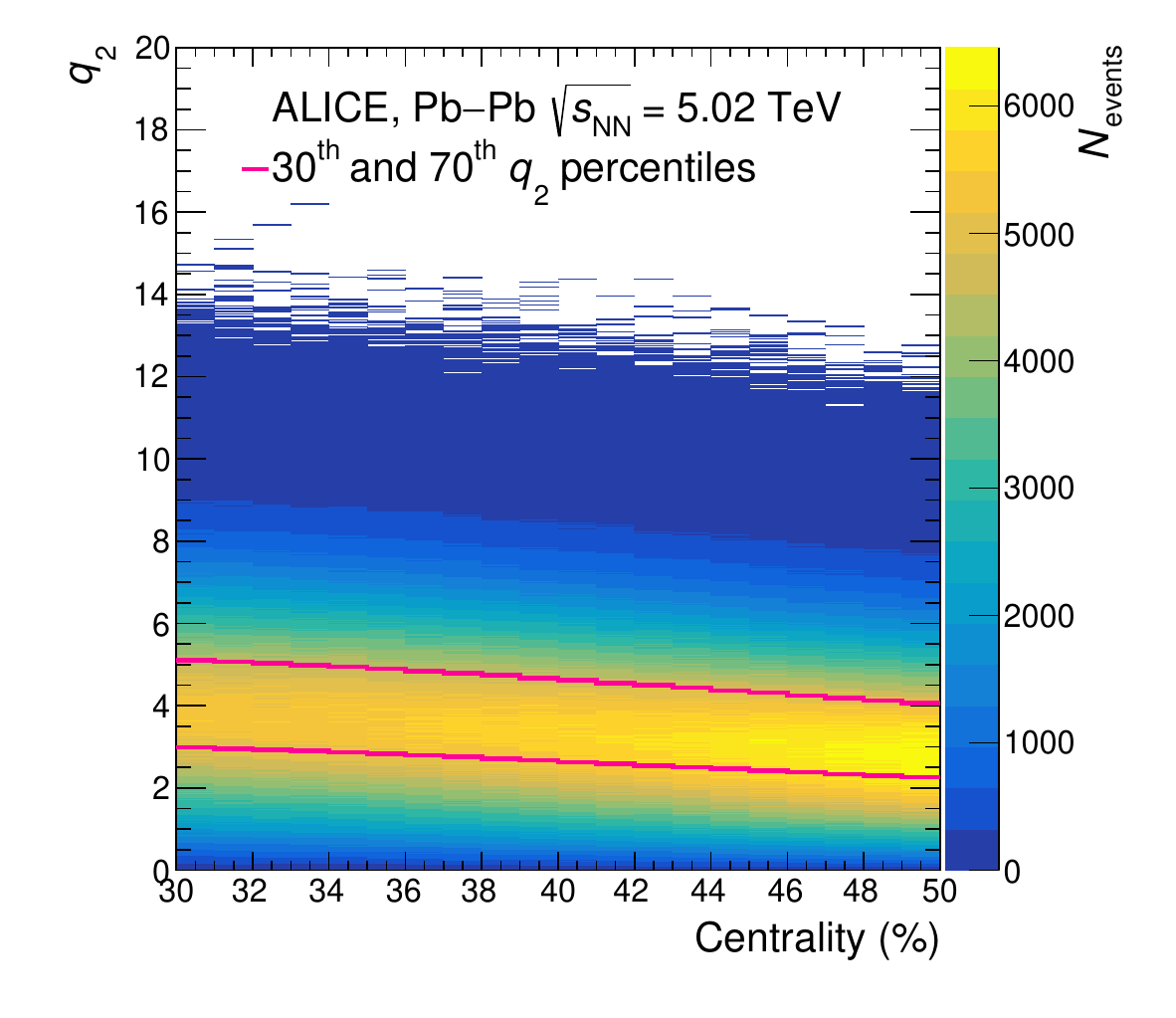}
    \end{center}
    \caption{Distribution of $q_{\rm 2}$ values as a function of centrality in \PbPb collisions at \fivenn. Pink lines demarcate the 30\textsuperscript{th} and 70\textsuperscript{th} $q_{\rm 2}$ percentiles, as calculated within $1\%$-wide centrality intervals.}
    \label{fig:q2}
\end{figure}

To study the event-shape dependence, events were classified according to the magnitude of the reduced flow vector $q_{\rm 2}$~\cite{STAR:2002hbo} as measured with the V0C, defined as
\begin{equation}
 q_{\rm 2} = |\textbf{Q}_{\rm 2}|/\sqrt{M},\label{eq:q2}
\end{equation}
where M represents the charged-particle multiplicity and \textbf{Q}\textsubscript{2} represents the second harmonic flow vector, defined as
\begin{equation}
        \textbf{Q}_{2} = \left( \sum_{i} w_{i} \cos(2\varphi_{i}), \sum_{i} w_{i} \sin(2\varphi_{i}) \right).
\end{equation}
Here, $\varphi_{i}$ and $w_{i}$ are the azimuthal angle and signal weight, respectively, of the $i$-th segment of the V0C detector~\cite{Voloshin:1994mz,ALICE:2014sbx}. Samples of events with the $30\%$ smallest and largest $q_{\rm 2}$ were selected for this study and will be henceforth referred to as $q_{\rm{2}}$-small and $q_{\rm{2}}$-large. These designations represent isotropic and elliptical event topologies, respectively. Figure~\ref{fig:q2} shows the distribution of $q_{\rm{2}}$ values as a function of collision centrality in \PbPb collisions at \fivenn. The pink lines demarcate the 30\textsuperscript{th} and 70\textsuperscript{th} percentiles in $q_{\rm{2}}$. Note that the $q_{\rm 2}$-small sample contains a significant fraction of events with non-zero $v_{\rm 2}$, so, while this sample is characterized as comparatively isotropic, there still exists some significant anisotropy within the sample~\cite{CF_ESE}. The slope of the distribution indicates that the average values of $q_{\rm 2}$ are slightly larger for more central collisions, which can introduce a centrality bias in the event class selection. To avoid this correlation bias, the $q_{\rm{2}}$ classification was done within $1\%$-wide centrality intervals.

The event-plane angle $\Psi_{2}$, given by the direction of $\textbf{Q}_{2}$, was measured with the V0A detector. The in- and out-of-plane axes were defined as parallel and perpendicular to $\Psi_{2}$, respectively. Jets were considered in- and out-of-plane when they were reconstructed within $30^{\rm o}$ in azimuth of these axes. This restriction from the traditional $\pm45^{\rm o}$ definition was made to enforce larger differences between in- and out-of-plane path lengths and to increase the potential differences in jet yields~\cite{Beattie:2022ojg}. The use of opposed detectors for  $q_{\rm 2}$ and $\Psi_{2}$ is advantageous for avoiding autocorrelations between these observables and for reducing detector-resolution corrections.

To account for the smearing of the reaction-plane angle due to the event-plane resolution, the ratios of in- and out-of-plane jet yields were corrected using a procedure analogous to that used for $v_{2}$ measurements~\cite{eventplane}. First, the $v_{2}$ was calculated using
\begin{equation}
    v_{\rm 2} = \frac{\pi}{3\sqrt{3}}\frac{1}{R_{\rm 2}}\frac{N_{\rm in} - N_{\rm out}}{N_{\rm in}+N_{\rm out}},
\label{eq:ResolutionCorrection}
\end{equation}
where $R_{\rm 2}$ is the second harmonic event-plane resolution, and $N_{\rm in}$ and $N_{\rm out}$ are the in- and out-of-plane jet yields, respectively. Note that the coefficient $\pi/(3\sqrt{3})$ in this formula is specific to this analysis, in which the in- and out-of-plane definitions are at $\pm30^{\rm o}$ around $\Psi_{2}$ and the vector perpendicular to it, as described above. The event-plane resolution $R_{\rm 2}$ was calculated using the three-sub-event method~\cite{eventplane}, where the particles measured by the V0A, V0C, and TPC detectors were used to construct the three separate sub-events. For $q_{\rm 2}$-small samples, $R_{\rm 2}$ is 0.55, whereas for $q_{\rm 2}$-large samples it is 0.68. After calculating the corrected $v_{\rm 2}$, the corrected ratio $\mathcal{R} = N_{\rm out}/N_{\rm in}$ was obtained by inverting Eq.~\ref{eq:ResolutionCorrection} and assuming a perfect resolution $R_{2}=1$. To correct the individual spectra for the event-plane resolution, conservation of jet yields within the fiducial volume ($N_{\rm in}^{\rm measured} + N_{\rm out}^{\rm measured} =  N_{\rm in}^{\rm corrected} + N_{\rm out}^{\rm corrected}$) was additionally considered, such that

\begin{equation}
    N_{\mathrm{in}}^{\mathrm{corrected}} = \frac{ N_{\mathrm{in}}^{\mathrm{measured}} + N_{\mathrm{out}}^{\mathrm{measured}}   }{  1 + \mathcal{R}   } ,
    \quad
    N_{\mathrm{out}}^{\mathrm{corrected}} = \frac{ N_{\mathrm{in}}^{\mathrm{measured}} + N_{\mathrm{out}}^{\mathrm{measured}}   }{  1 + 1 / \mathcal{R}  }.
\label{eq:ResolutionCorrection2}
\end{equation}

For the ratio of out-of-plane to in-plane jet yields, the magnitude of this correction varies from 5 to 25\%. Note that $N_{\rm mid}$ remains unmodified, where $N_{\rm mid}$ is the jet yield reconstructed between $\pm30^{\rm o}-60^{\rm o}$ of the event plane. This correction procedure is exact when assuming a negligible contribution from  higher order harmonics. Additionally, the contribution of non-flow to the measured yield ratios was estimated using PYTHIA 8. Here, non-flow refers to the $v_{\rm 2}$ contribution from forward multi-jets that result in a biased determination of $\Psi_{\rm 2}$. It was found that, for cases where an intermediate \jetpt jet is produced at midrapidity, a recoiling jet strikes the V0A in $<4\%$ of instances. The relative contribution from these events to the jet $v_{2}$ is estimated to be less than $20\%$. The presented results are not corrected for this possible effect.

The systematic uncertainties of the charged-particle jet yields and their ratios are summarized in Tables~\ref{Table:systematics} and~\ref{Table:systematicsRatio}, respectively. The ranges of systematic uncertainties are listed for the measured $\jetpt$ range. The systematic uncertainty on the tracking efficiency was calculated by randomly rejecting an additional $4\%$ of PYTHIA 8 tracks used in the embedding procedure, representing the uncertainty in the single-track efficiency in the Pb--Pb environment. The jet finding was then repeated and the response matrix recalculated, resulting in the largest source of uncertainty for the measured spectra. The uncertainty in the unfolding procedure was quantified by varying the number of iterations of unfolding, the shape of the prior $\jetpt$ and $\Delta\varphi$ spectra, and the lower limit of the measured range (referred to as the truncation). The shape variation was done by reweighting the unfolding prior according to the ratio between the PYTHIA 8 and data spectra in both \jetpt and event-plane angle. The number of unfolding iterations was varied by $\pm$1. The lower \jetpt limit for the jets that entered into the unfolding procedure was varied by $\pm5~{\rm GeV}/c$. Finally, the systematic uncertainty of the event-plane resolution was obtained by varying $R_{2}$ by 2\%. This $2\%$ variation accounts for the difference in event-plane resolution observed when it is calculated using the $\chi$-ratio method as opposed to the three-sub-event method~\cite{eventplane,CF_ESE}. Note that this uncertainty is only considered for the measurements that are differentiated in $\Delta\varphi$. For the ratios of the spectra, the systematic uncertainties in the numerator and denominator were treated as correlated, and the resulting systematic uncertainty was obtained by making the above-described variations and calculating the deviations on the ratio itself. The total systematic uncertainties were calculated as quadratic sums of the different sources by assuming the independence of all contributions.

\begin{table}
\centering
\caption{Relative systematic uncertainties for the charged-particle jet yields as measured in 30--50\% \PbPb collisions at \fivenn. Values are reported as percentages. Reported ranges reflect the minimum and maximum values of the uncertainties over the measured \jetpt range. Here, $<$ 1 indicates an uncertainty with a decimal value greater than zero but less than one.}
\scalebox{0.75}{
\begin{tabular}{l|cccc|cccc}
\cline{2-9}
     & \multicolumn{4}{c|}{\textbf{\textit{R} = 0.2}} & \multicolumn{4}{c}{\textbf{\textit{R} = 0.4}} \\
     & \multicolumn{2}{c}{$q_{\rm 2}$-small} & \multicolumn{2}{c|}{$q_{\rm 2}$-large}
     & \multicolumn{2}{c}{$q_{\rm 2}$-small} & \multicolumn{2}{c}{$q_{\rm 2}$-large}\\
     & in-plane & out-of-plane & in-plane & out-of-plane
     & in-plane & out-of-plane & in-plane & out-of-plane\\ [0.9ex]
\hline\hline
    Tracking efficiency         & 6--15 & 6--12 & 8--10 & 6--16     & 4--15 & 6--15 & 6--20 & $<$1--17       \\[1ex] 
    Unfolding iterations        & $<$1 & $<$1 & $<$1 & $<$1    & $<$1--2 & $<$1--4 & 1--3 & $<$1--3        \\[1ex] 
    Unfolding prior             & $<$1--3 & $<$1--2 & $<$1--2 & $<$1--2 & 2--6 & $<$1--5 & $<$1--8 & 2--5        \\[1ex] 
    Unfolding truncation        & $<$1 & $<$1 & $<$1 & $<$1    & $<$1--10 & $<$1--7 & 1--13 & $<$1--9        \\[1ex]
    Event-plane determination   & $<$1 & $<$1 & $<$1 & $<$1    & $<$1 & $<$1 & $<$1 & $<$1     \\[1ex] 
\hline
    \textbf{Total}              & 6--15 & 6--12 & 8--10 & 6--16     & 8--16 & 6--16 & 12--21 & 9--17       \\[1ex]
\hline
\end{tabular}
}
\label{Table:systematics}
\end{table}

\begin{table}
\centering
\caption{Relative systematic uncertainties for the ratios of charged-particle jet yields as measured in 30--50\% \PbPb collisions at \fivenn. Values are reported as percentages. Reported ranges reflect the minimum and maximum values of the uncertainties over the measured \jetpt range. Here, $<$ 1 indicates an uncertainty with a decimal value greater than zero but less than one.}
\scalebox{0.75}{
\begin{tabular}{l|ccc|ccc}
\cline{2-7}
     & \multicolumn{3}{c|}{\textbf{\textit{R} = 0.2}} & \multicolumn{3}{c}{\textbf{\textit{R} = 0.4}} \\
     & $q_{\rm 2}$-large/$q_{\rm 2}$-small & $q_{\rm 2}$-small & $q_{\rm 2}$-large
     & $q_{\rm 2}$-large/$q_{\rm 2}$-small & $q_{\rm 2}$-small & $q_{\rm 2}$-large \\ [0.9ex]
     & & out-/in-plane & out-/in-plane 
     & & out-/in-plane & out-/in-plane \\ [0.9ex]
\hline\hline
    Tracking efficiency         & 1--3 & $<$1--2 & $<$1--5    & 1--3 & 4--9 & 1--9      \\[1ex] 
    Unfolding iterations        & $<$1 & $<$1 & $<$1    & $<$1 & $<$1--6 & 2--6       \\[1ex] 
    Unfolding prior             & $<$1--3 & $<$1--2 & $<$1--3    & $<$1--3 & 1--5 & $<$1--12       \\[1ex] 
    Unfolding truncation        & $<$1 & $<$1 & $<$1    & $<$1--3 & 1--18 & 1--21        \\[1ex]
    Event-plane determination   & N/A & $<$1 & $<$1    & N/A & $<$1 & $<$1   \\[1ex] 
\hline
    \textbf{Total}              & 1--4 & 2--3 & 1--5    & 1--5 & 5--21 & 4--25        \\[1ex]
\hline
\end{tabular}
}
\label{Table:systematicsRatio}
\end{table}

\section{Results}

\begin{figure}[h!]
    \begin{center}
    \includegraphics[width = 0.48\textwidth]{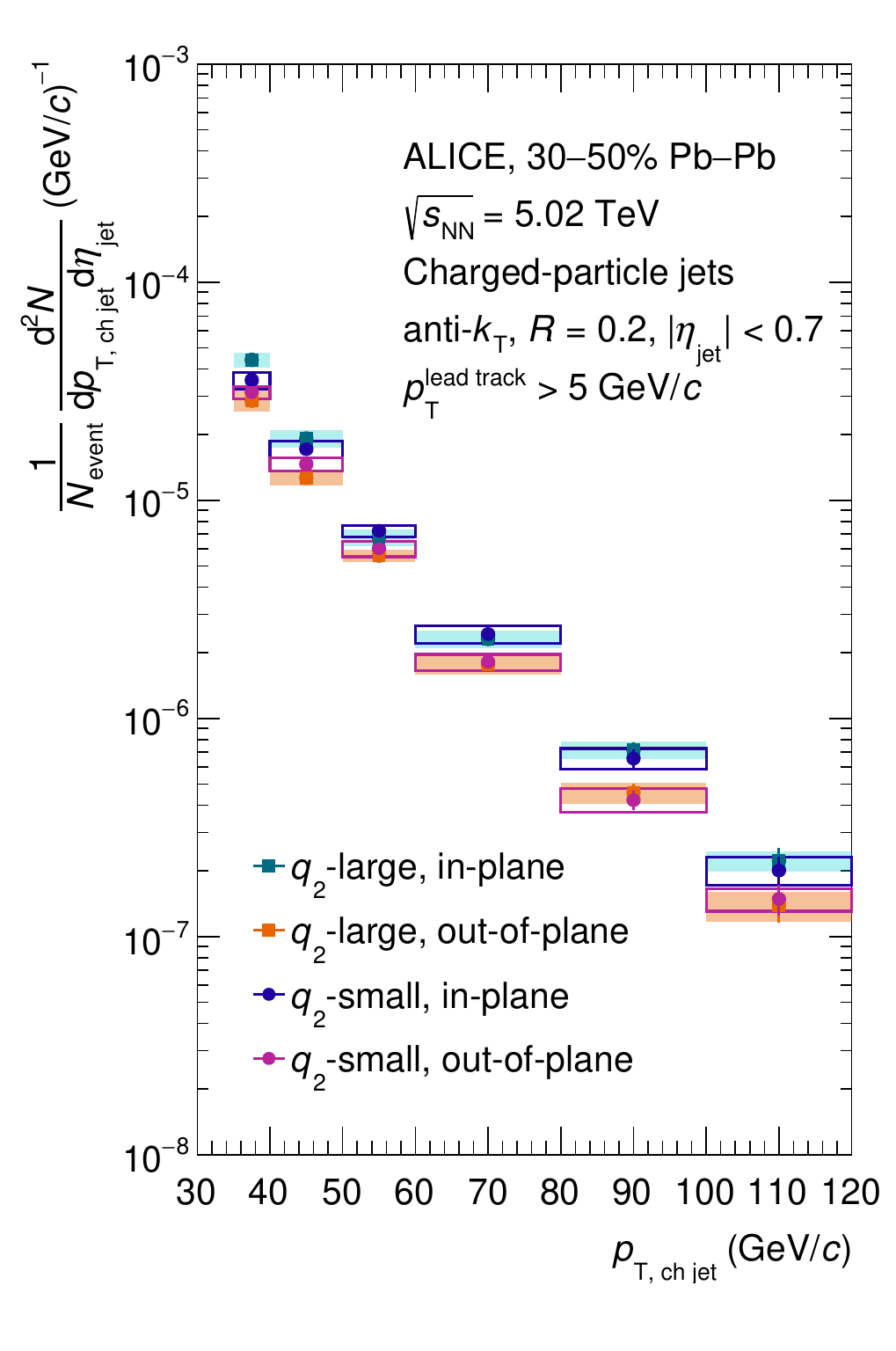}
    \includegraphics[width = 0.48\textwidth]{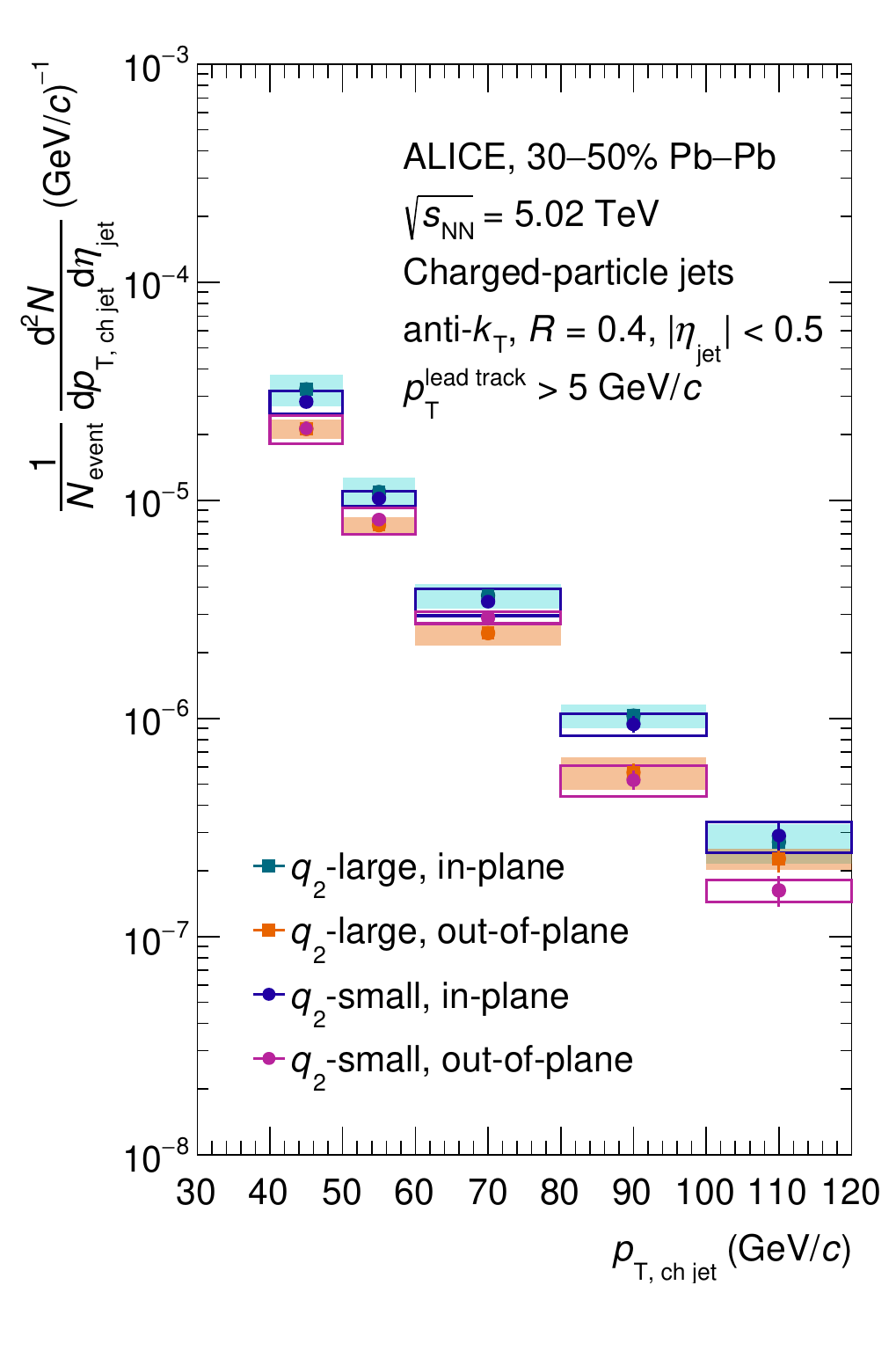}
    \end{center}
    \caption{Charged-particle jet yields for $R=0.2$ (left) and $R=0.4$ (right) jets in \PbPb collisions at \fivenn. Results are shown for the $q_{\rm 2}$-small and $q_{\rm 2}$-large event classes, for in- and out-of-plane jets. The bars (boxes) represent statistical (systematic) point-by-point uncertainties.}
    \label{fig:ESEJetSpectra}
\end{figure}

The \jetpt-differential charged-particle jet yields for resolution parameters $R=0.2$ and $R=0.4$ are shown in Fig.~\ref{fig:ESEJetSpectra}. Included are the results for the event classes $q_{\rm 2}$-small and $q_{\rm 2}$-large, differentiated for in-plane and out-of-plane jets. The systematic uncertainties are indicated by the boxes and are highly correlated among the different measurements.

\begin{figure}[tb]
    \begin{center}
    \includegraphics[width = 0.6\textwidth]{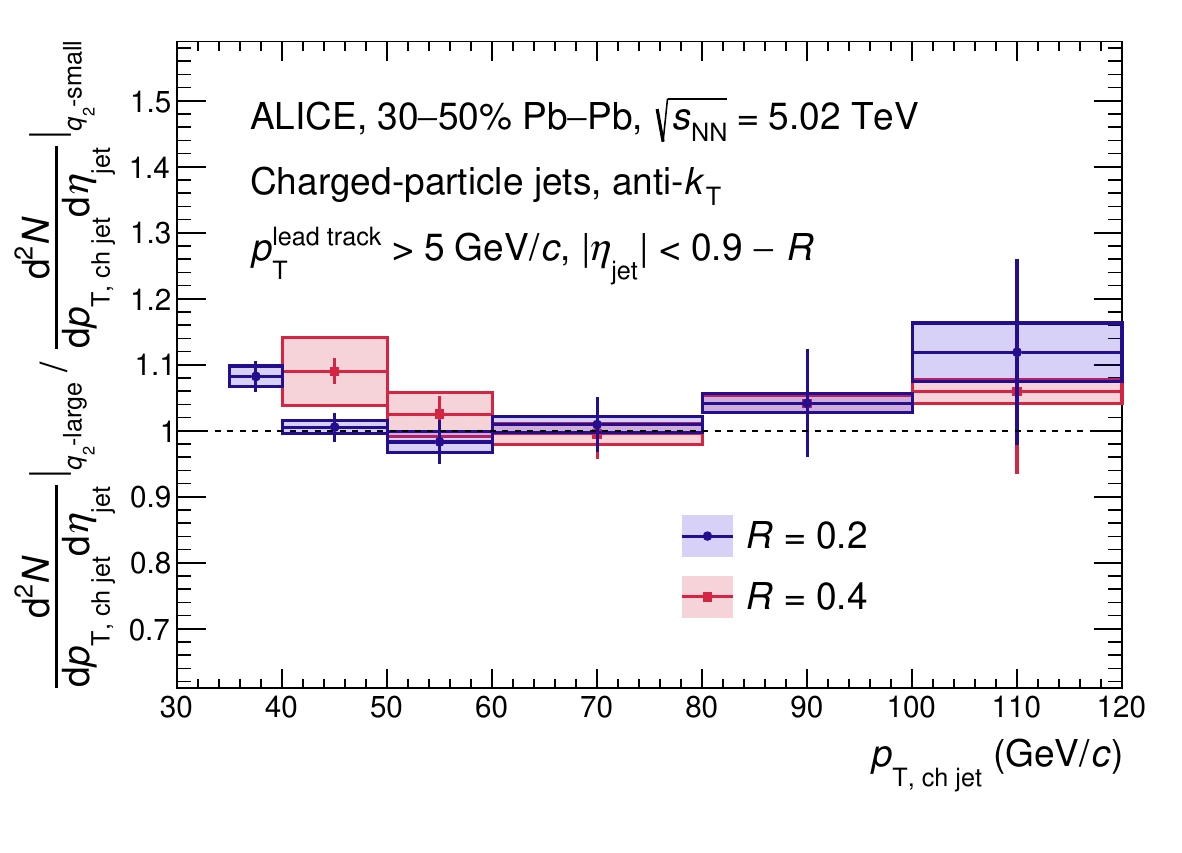}
    \end{center}
    \caption{Ratio of the charged-particle jet yields of the $q_{\rm 2}$-large to the $q_{\rm 2}$-small event classes, in \PbPb collisions at \fivenn. The results are reported for $R = 0.2$ and $R = 0.4$ jets.}
    \label{fig:ESEratios}
\end{figure}

The ratios of charged-particle jet yields for the $q_{\rm 2}$-large to $q_{\rm 2}$-small event classes are shown in Fig.~\ref{fig:ESEratios}, for both $R=0.2$ and $R=0.4$. Considering these results as ratios allowed for a reduction in the systematic uncertainties due to correlations of the uncertainties between the spectra, thus improving the sensitivity of this measurement. The results are consistent with unity, indicating that azimuthally-integrated jet yields are not sensitive to collision ellipticity. This stands in contrast to the yield enhancement seen in elliptical collisions at low $p_{\rm T}$~\cite{ESE276}, where particle spectra are not governed by quenching, but rather by the hydrodynamic expansion of the medium.

\begin{figure}[tb]
    \begin{center}
    \includegraphics[width = 0.49\textwidth]{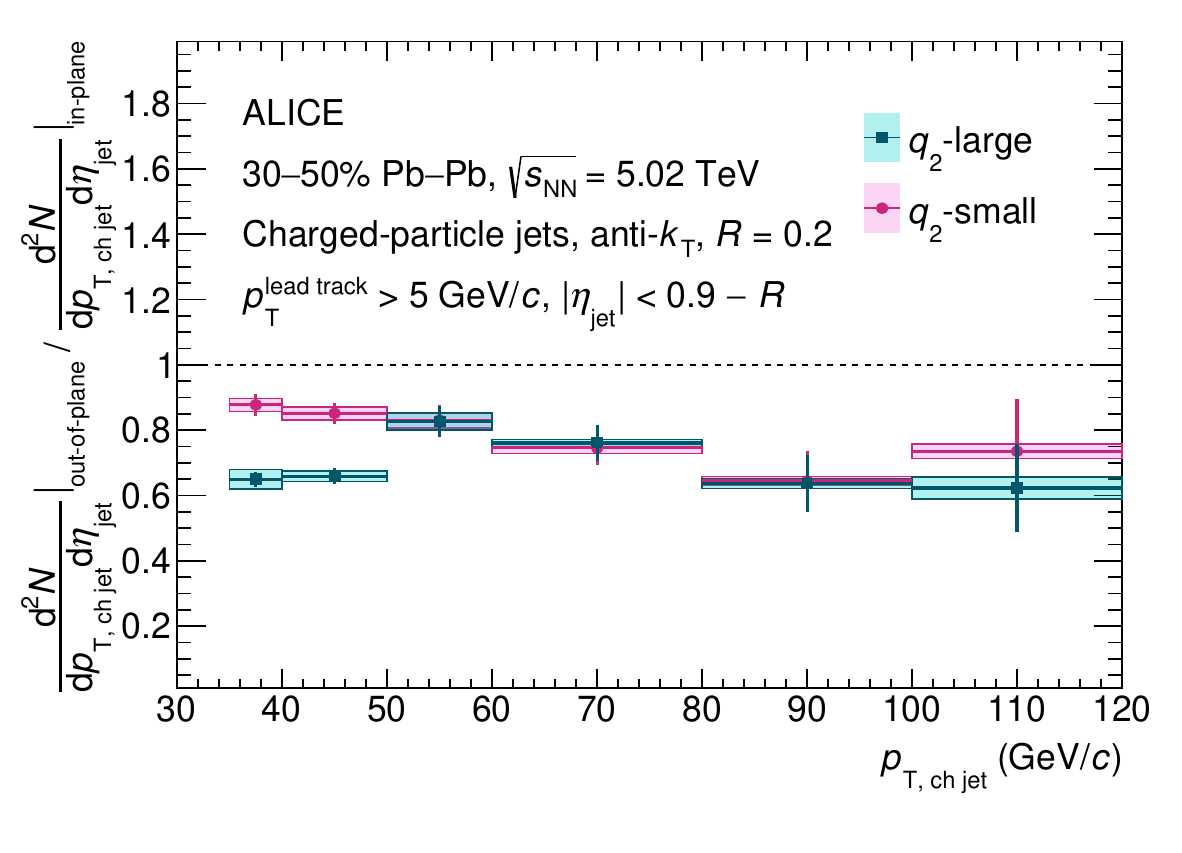}
    \includegraphics[width = 0.49\textwidth]{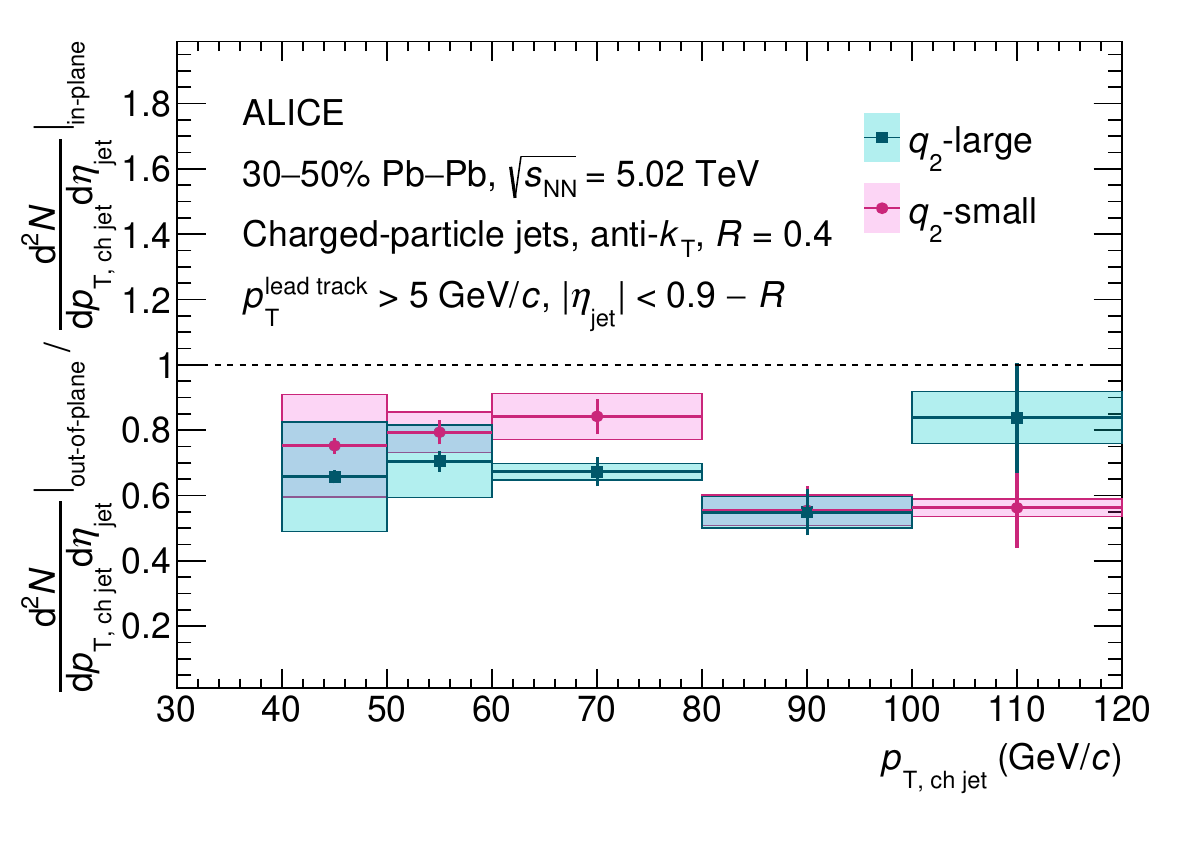}
    \end{center}
    \caption{Ratios of out-of-plane to in-plane charged-particle jet yields for the $q_{\rm 2}$-large and $q_{\rm 2}$-small event classes in \PbPb collisions at \fivenn. The results are reported for $R = 0.2$ (left) and $R = 0.4$ (right) jets, and are corrected for the event-plane resolution.}
    \label{fig:CorrectedRatios}
\end{figure}

Figure~\ref{fig:CorrectedRatios} shows the ratio of out-of-plane to in-plane jet yields for the $q_{\rm 2}$-small and $q_{\rm 2}$-large event classes, for jets with $R=0.2$ (left) and $R=0.4$ (right). These results were corrected for the event-plane resolution, as described in the previous section. The measured ratios are significantly below unity, indicating that jets lose more energy on average when traveling out-of-plane than when traveling in-plane. This is consistent with the idea that the magnitude of jet energy loss is driven, at least in part, by the path length traversed in the medium. For $R=0.4$ jets, further conclusions regarding event-shape dependent azimuthal anisotropy are limited by the large experimental uncertainties. For $R=0.2$ jets, the ratios for $q_{\rm 2}$-small and $q_{\rm 2}$-large are similar at high \jetpt. For $\jetpt<50$ GeV/\textit{c}, however, there is an indication that out-of-plane jets in the $q_{\rm 2}$-large class are more suppressed relative to in-plane jets than those in the $q_{\rm 2}$-small class. The significance of this separation from $35 < \jetpt<50$ GeV/\textit{c} is 5.2$\sigma$. This result is qualitatively in agreement with observations of D mesons~\cite{HF2}.\par

This effect is expected due to the increased path-length differences between in- and out-of-plane directions for highly elliptical collision geometries, which is supported by Trajectum calculations~\cite{trajectum, Beattie:2022ojg}. In these calculations, probes were generated in the initial state at the location of nucleon--nucleon collisions, and propagated through the hydrodynamically evolving medium while remaining unmodified. The average path lengths of these probes were calculated for events with varying  $q_{\rm 2}$, and differentially for in- and out-of-plane emission angles. While the results of this study show that the average traversed path length of the probes does not vary significantly with event $q_{\rm 2}$, it does change as a function of the probe angle with respect to $\Psi_{\rm 2}$. This variation with $\Psi_{\rm 2}$ can be further augmented when considering $q_{\rm 2}$-large events, and suppressed when considering $q_{\rm 2}$-small events. The outcome of these Trajectum calculations shows that by using ESE, the ratio of out-of-plane to in-plane path lengths can be increased in semicentral collisions by $\sim$10\% with respect to the inclusive case. The results presented in this letter are consistent with these calculations, assuming that the traversed path length of jets is an important factor for determining their energy loss. These Trajectum studies do not, however, allow one to conclude anything about the \jetpt-dependence of this energy loss or explain the apparent convergence of ratios at high \jetpt. Despite the absence of phenomenological descriptions, a possible understanding of the experimental results at high \jetpt can be obtained by considering that the charged-particle jet $R_{\rm AA}$ increases and the charged-particle jet $v_{\rm 2}$ decreases with increasing \jetpt~\cite{jetv2, ALICEchargedjetRaa}. It is therefore expected that any path-length-dependent signal accessible to ESE measurements would decrease at high \jetpt. Moreover, the precision of the measurement presented here is statistically limited at high \jetpt. It is therefore difficult to establish if the convergence of out-of-plane to in-plane ratios for elliptical and isotropic events is a true physics phenomenon, or is rather a consequence of the limited experimental precision accessible at high \jetpt.

This measurement demonstrates the potential of the ESE technique and paves the way for future studies with larger data samples. However, a full interpretation of these results requires detailed comparisons to model calculations, which will allow for more quantitative conclusions on the path-length dependence of energy loss.

\section{Conclusions}

In this letter, the measured event-shape engineered jet yields are reported for resolution parameters $R=0.2$ and $R=0.4$ in semicentral \PbPb collisions at \fivenn. The magnitude of the reduced second harmonic flow vector $q_{\rm 2}$ was used to select event classes that are particularly isotropic ($q_{\rm 2}$-small) and elliptical ($q_{\rm 2}$-large). The jet spectra from these two event classes are consistent within their uncertainties. However, a significant deviation between jet spectra is observed when these jets are classified according to their azimuthal angle with respect to the event plane. It is indicated that jets lose more energy out-of-plane compared to in-plane, consistent with the measurement of a non-zero $v_{2}$ of jets at the LHC. Furthermore, for $R=0.2$ jets in highly elliptical events, the differences between the modification of out-of-plane and in-plane jets at low \jetpt are found to be more significant than in more isotropic events. Model calculations employing  a realistic parton shower in event-by-event hydrodynamical simulations, such as LBT~\cite{He:2015pra, He:2018xjv}, JETSCAPE~\cite{Putschke:2019yrg}, or JEWEL on a (2+1)D background~\cite{Barreto:2022ulg,Kolbe:2023rsq}, are needed to further interpret these results and gain insight into the path-length dependence of jet quenching.

%%%%%%%%%%%%%%%%%%%%%%%%%%%%%%%%
% end main text 
%%%%%%%%%%%%%%%%%%%%%%%%%%%%%%%%

%%%%% acknowledgements - handled by EB chairs 
\newenvironment{acknowledgement}{\relax}{\relax}
\begin{acknowledgement}
\section*{Acknowledgements}
% add specific acknowledgements here 
% ...but please don't remove the line below: funding agencies
% will be acknowledged with a custom tex file handled by EB chairs after Collab Round 2
% Version: 2023-07-07

The ALICE Collaboration would like to thank all its engineers and technicians for their invaluable contributions to the construction of the experiment and the CERN accelerator teams for the outstanding performance of the LHC complex.
The ALICE Collaboration gratefully acknowledges the resources and support provided by all Grid centres and the Worldwide LHC Computing Grid (WLCG) collaboration.
The ALICE Collaboration acknowledges the following funding agencies for their support in building and running the ALICE detector:
A. I. Alikhanyan National Science Laboratory (Yerevan Physics Institute) Foundation (ANSL), State Committee of Science and World Federation of Scientists (WFS), Armenia;
Austrian Academy of Sciences, Austrian Science Fund (FWF): [M 2467-N36] and Nationalstiftung f\"{u}r Forschung, Technologie und Entwicklung, Austria;
Ministry of Communications and High Technologies, National Nuclear Research Center, Azerbaijan;
Conselho Nacional de Desenvolvimento Cient\'{\i}fico e Tecnol\'{o}gico (CNPq), Financiadora de Estudos e Projetos (Finep), Funda\c{c}\~{a}o de Amparo \`{a} Pesquisa do Estado de S\~{a}o Paulo (FAPESP) and Universidade Federal do Rio Grande do Sul (UFRGS), Brazil;
Bulgarian Ministry of Education and Science, within the National Roadmap for Research Infrastructures 2020-2027 (object CERN), Bulgaria;
Ministry of Education of China (MOEC) , Ministry of Science \& Technology of China (MSTC) and National Natural Science Foundation of China (NSFC), China;
Ministry of Science and Education and Croatian Science Foundation, Croatia;
Centro de Aplicaciones Tecnol\'{o}gicas y Desarrollo Nuclear (CEADEN), Cubaenerg\'{\i}a, Cuba;
Ministry of Education, Youth and Sports of the Czech Republic, Czech Republic;
The Danish Council for Independent Research | Natural Sciences, the VILLUM FONDEN and Danish National Research Foundation (DNRF), Denmark;
Helsinki Institute of Physics (HIP), Finland;
Commissariat \`{a} l'Energie Atomique (CEA) and Institut National de Physique Nucl\'{e}aire et de Physique des Particules (IN2P3) and Centre National de la Recherche Scientifique (CNRS), France;
Bundesministerium f\"{u}r Bildung und Forschung (BMBF) and GSI Helmholtzzentrum f\"{u}r Schwerionenforschung GmbH, Germany;
General Secretariat for Research and Technology, Ministry of Education, Research and Religions, Greece;
National Research, Development and Innovation Office, Hungary;
Department of Atomic Energy Government of India (DAE), Department of Science and Technology, Government of India (DST), University Grants Commission, Government of India (UGC) and Council of Scientific and Industrial Research (CSIR), India;
National Research and Innovation Agency - BRIN, Indonesia;
Istituto Nazionale di Fisica Nucleare (INFN), Italy;
Japanese Ministry of Education, Culture, Sports, Science and Technology (MEXT) and Japan Society for the Promotion of Science (JSPS) KAKENHI, Japan;
Consejo Nacional de Ciencia (CONACYT) y Tecnolog\'{i}a, through Fondo de Cooperaci\'{o}n Internacional en Ciencia y Tecnolog\'{i}a (FONCICYT) and Direcci\'{o}n General de Asuntos del Personal Academico (DGAPA), Mexico;
Nederlandse Organisatie voor Wetenschappelijk Onderzoek (NWO), Netherlands;
The Research Council of Norway, Norway;
Commission on Science and Technology for Sustainable Development in the South (COMSATS), Pakistan;
Pontificia Universidad Cat\'{o}lica del Per\'{u}, Peru;
Ministry of Education and Science, National Science Centre and WUT ID-UB, Poland;
Korea Institute of Science and Technology Information and National Research Foundation of Korea (NRF), Republic of Korea;
Ministry of Education and Scientific Research, Institute of Atomic Physics, Ministry of Research and Innovation and Institute of Atomic Physics and Universitatea Nationala de Stiinta si Tehnologie Politehnica Bucuresti, Romania;
Ministry of Education, Science, Research and Sport of the Slovak Republic, Slovakia;
National Research Foundation of South Africa, South Africa;
Swedish Research Council (VR) and Knut \& Alice Wallenberg Foundation (KAW), Sweden;
European Organization for Nuclear Research, Switzerland;
Suranaree University of Technology (SUT), National Science and Technology Development Agency (NSTDA) and National Science, Research and Innovation Fund (NSRF via PMU-B B05F650021), Thailand;
Turkish Energy, Nuclear and Mineral Research Agency (TENMAK), Turkey;
National Academy of  Sciences of Ukraine, Ukraine;
Science and Technology Facilities Council (STFC), United Kingdom;
National Science Foundation of the United States of America (NSF) and United States Department of Energy, Office of Nuclear Physics (DOE NP), United States of America.
In addition, individual groups or members have received support from:
European Research Council, Strong 2020 - Horizon 2020 (grant nos. 950692, 824093), European Union;
Academy of Finland (Center of Excellence in Quark Matter) (grant nos. 346327, 346328), Finland.

\end{acknowledgement}

%%%%%%%% Bibliography 
\bibliographystyle{utphys}   % Remember we use title in the biblio
\bibliography{bibliography}
%\input {bibliography.tex}  

%%%%%%%%%%%%%%%%%%%%%%%%%%%%%%%%
% Appendices: yours (if any) + authorlist
%%%%%%%%%%%%%%%%%%%%%%%%%%%%%%%%
\newpage
\appendix

%
%\input{} % put your appendices here (if any)
%

%%%%% Authorlist - please do not touch: handled by EB chairs 
\section{The ALICE Collaboration}
\label{app:collab}
% ALICE Collaboration author list for 2023-07-07
\begin{flushleft} 
\small

S.~Acharya\,\orcidlink{0000-0002-9213-5329}\,$^{\rm 128}$, 
D.~Adamov\'{a}\,\orcidlink{0000-0002-0504-7428}\,$^{\rm 87}$, 
G.~Aglieri Rinella\,\orcidlink{0000-0002-9611-3696}\,$^{\rm 33}$, 
M.~Agnello\,\orcidlink{0000-0002-0760-5075}\,$^{\rm 30}$, 
N.~Agrawal\,\orcidlink{0000-0003-0348-9836}\,$^{\rm 52}$, 
Z.~Ahammed\,\orcidlink{0000-0001-5241-7412}\,$^{\rm 136}$, 
S.~Ahmad\,\orcidlink{0000-0003-0497-5705}\,$^{\rm 16}$, 
S.U.~Ahn\,\orcidlink{0000-0001-8847-489X}\,$^{\rm 72}$, 
I.~Ahuja\,\orcidlink{0000-0002-4417-1392}\,$^{\rm 38}$, 
A.~Akindinov\,\orcidlink{0000-0002-7388-3022}\,$^{\rm 142}$
% ...

\section*{Affiliation Notes}

$^{\rm I}$ Deceased\\
$^{\rm II}$ Also at: Max-Planck-Institut fur Physik, Munich, Germany\\
$^{\rm III}$ Also at: Italian National Agency for New Technologies, Energy and Sustainable Economic Development (ENEA), Bologna, Italy\\
$^{\rm IV}$ Also at: Dipartimento DET del Politecnico di Torino, Turin, Italy\\
$^{\rm V}$ Also at: Yildiz Technical University, Istanbul, T\"{u}rkiye\\
$^{\rm VI}$ Also at: Department of Applied Physics, Aligarh Muslim University, Aligarh, India\\
$^{\rm VII}$ Also at: Institute of Theoretical Physics, University of Wroclaw, Poland\\
$^{\rm VIII}$ Also at: An institution covered by a cooperation agreement with CERN\\

\section*{Collaboration Institutes}

$^{1}$ A.I. Alikhanyan National Science Laboratory (Yerevan Physics Institute) Foundation, Yerevan, Armenia\\
$^{2}$ AGH University of Krakow, Cracow, Poland\\
$^{3}$ Bogolyubov Institute for Theoretical Physics, National Academy of Sciences of Ukraine, Kiev, Ukraine\\
$^{4}$ Bose Institute, Department of Physics  and Centre for Astroparticle Physics and Space Science (CAPSS), Kolkata, India\\
$^{5}$ California Polytechnic State University, San Luis Obispo, California, United States\\
$^{6}$ Central China Normal University, Wuhan, China\\
$^{7}$ Centro de Aplicaciones Tecnol\'{o}gicas y Desarrollo Nuclear (CEADEN), Havana, Cuba\\
$^{8}$ Centro de Investigaci\'{o}n y de Estudios Avanzados (CINVESTAV), Mexico City and M\'{e}rida, Mexico\\
$^{9}$ Chicago State University, Chicago, Illinois, United States\\
$^{10}$ China Institute of Atomic Energy, Beijing, China\\
$^{11}$ China University of Geosciences, Wuhan, China\\
$^{12}$ Chungbuk National University, Cheongju, Republic of Korea\\
$^{13}$ Comenius University Bratislava, Faculty of Mathematics, Physics and Informatics, Bratislava, Slovak Republic\\
$^{14}$ COMSATS University Islamabad, Islamabad, Pakistan\\
$^{15}$ Creighton University, Omaha, Nebraska, United States\\
$^{16}$ Department of Physics, Aligarh Muslim University, Aligarh, India\\
$^{17}$ Department of Physics, Pusan National University, Pusan, Republic of Korea\\
$^{18}$ Department of Physics, Sejong University, Seoul, Republic of Korea\\
$^{19}$ Department of Physics, University of California, Berkeley, California, United States\\
$^{20}$ Department of Physics, University of Oslo, Oslo, Norway\\
$^{21}$ Department of Physics and Technology, University of Bergen, Bergen, Norway\\
$^{22}$ Dipartimento di Fisica, Universit\`{a} di Pavia, Pavia, Italy\\
$^{23}$ Dipartimento di Fisica dell'Universit\`{a} and Sezione INFN, Cagliari, Italy\\
$^{24}$ Dipartimento di Fisica dell'Universit\`{a} and Sezione INFN, Trieste, Italy\\
$^{25}$ Dipartimento di Fisica dell'Universit\`{a} and Sezione INFN, Turin, Italy\\
$^{26}$ Dipartimento di Fisica e Astronomia dell'Universit\`{a} and Sezione INFN, Bologna, Italy\\
$^{27}$ Dipartimento di Fisica e Astronomia dell'Universit\`{a} and Sezione INFN, Catania, Italy\\
$^{28}$ Dipartimento di Fisica e Astronomia dell'Universit\`{a} and Sezione INFN, Padova, Italy\\
$^{29}$ Dipartimento di Fisica `E.R.~Caianiello' dell'Universit\`{a} and Gruppo Collegato INFN, Salerno, Italy\\
$^{30}$ Dipartimento DISAT del Politecnico and Sezione INFN, Turin, Italy\\
$^{31}$ Dipartimento di Scienze MIFT, Universit\`{a} di Messina, Messina, Italy\\
$^{32}$ Dipartimento Interateneo di Fisica `M.~Merlin' and Sezione INFN, Bari, Italy\\
$^{33}$ European Organization for Nuclear Research (CERN), Geneva, Switzerland\\
$^{34}$ Faculty of Electrical Engineering, Mechanical Engineering and Naval Architecture, University of Split, Split, Croatia\\
$^{35}$ Faculty of Engineering and Science, Western Norway University of Applied Sciences, Bergen, Norway\\
$^{36}$ Faculty of Nuclear Sciences and Physical Engineering, Czech Technical University in Prague, Prague, Czech Republic\\
$^{37}$ Faculty of Physics, Sofia University, Sofia, Bulgaria\\
$^{38}$ Faculty of Science, P.J.~\v{S}af\'{a}rik University, Ko\v{s}ice, Slovak Republic\\
$^{39}$ Frankfurt Institute for Advanced Studies, Johann Wolfgang Goethe-Universit\"{a}t Frankfurt, Frankfurt, Germany\\
$^{40}$ Fudan University, Shanghai, China\\
$^{41}$ Gangneung-Wonju National University, Gangneung, Republic of Korea\\
$^{42}$ Gauhati University, Department of Physics, Guwahati, India\\
$^{43}$ Helmholtz-Institut f\"{u}r Strahlen- und Kernphysik, Rheinische Friedrich-Wilhelms-Universit\"{a}t Bonn, Bonn, Germany\\
$^{44}$ Helsinki Institute of Physics (HIP), Helsinki, Finland\\
$^{45}$ High Energy Physics Group,  Universidad Aut\'{o}noma de Puebla, Puebla, Mexico\\
$^{46}$ Horia Hulubei National Institute of Physics and Nuclear Engineering, Bucharest, Romania\\
$^{47}$ HUN-REN Wigner Research Centre for Physics, Budapest, Hungary\\
$^{48}$ Indian Institute of Technology Bombay (IIT), Mumbai, India\\
$^{49}$ Indian Institute of Technology Indore, Indore, India\\
$^{50}$ INFN, Laboratori Nazionali di Frascati, Frascati, Italy\\
$^{51}$ INFN, Sezione di Bari, Bari, Italy\\
$^{52}$ INFN, Sezione di Bologna, Bologna, Italy\\
$^{53}$ INFN, Sezione di Cagliari, Cagliari, Italy\\
$^{54}$ INFN, Sezione di Catania, Catania, Italy\\
$^{55}$ INFN, Sezione di Padova, Padova, Italy\\
$^{56}$ INFN, Sezione di Pavia, Pavia, Italy\\
$^{57}$ INFN, Sezione di Torino, Turin, Italy\\
$^{58}$ INFN, Sezione di Trieste, Trieste, Italy\\
$^{59}$ Inha University, Incheon, Republic of Korea\\
$^{60}$ Institute for Gravitational and Subatomic Physics (GRASP), Utrecht University/Nikhef, Utrecht, Netherlands\\
$^{61}$ Institute of Experimental Physics, Slovak Academy of Sciences, Ko\v{s}ice, Slovak Republic\\
$^{62}$ Institute of Physics, Homi Bhabha National Institute, Bhubaneswar, India\\
$^{63}$ Institute of Physics of the Czech Academy of Sciences, Prague, Czech Republic\\
$^{64}$ Institute of Space Science (ISS), Bucharest, Romania\\
$^{65}$ Institut f\"{u}r Kernphysik, Johann Wolfgang Goethe-Universit\"{a}t Frankfurt, Frankfurt, Germany\\
$^{66}$ Instituto de Ciencias Nucleares, Universidad Nacional Aut\'{o}noma de M\'{e}xico, Mexico City, Mexico\\
$^{67}$ Instituto de F\'{i}sica, Universidade Federal do Rio Grande do Sul (UFRGS), Porto Alegre, Brazil\\
$^{68}$ Instituto de F\'{\i}sica, Universidad Nacional Aut\'{o}noma de M\'{e}xico, Mexico City, Mexico\\
$^{69}$ iThemba LABS, National Research Foundation, Somerset West, South Africa\\
$^{70}$ Jeonbuk National University, Jeonju, Republic of Korea\\
$^{71}$ Johann-Wolfgang-Goethe Universit\"{a}t Frankfurt Institut f\"{u}r Informatik, Fachbereich Informatik und Mathematik, Frankfurt, Germany\\
$^{72}$ Korea Institute of Science and Technology Information, Daejeon, Republic of Korea\\
$^{73}$ KTO Karatay University, Konya, Turkey\\
$^{74}$ Laboratoire de Physique Subatomique et de Cosmologie, Universit\'{e} Grenoble-Alpes, CNRS-IN2P3, Grenoble, France\\
$^{75}$ Lawrence Berkeley National Laboratory, Berkeley, California, United States\\
$^{76}$ Lund University Department of Physics, Division of Particle Physics, Lund, Sweden\\
$^{77}$ Nagasaki Institute of Applied Science, Nagasaki, Japan\\
$^{78}$ Nara Women{'}s University (NWU), Nara, Japan\\
$^{79}$ National and Kapodistrian University of Athens, School of Science, Department of Physics , Athens, Greece\\
$^{80}$ National Centre for Nuclear Research, Warsaw, Poland\\
$^{81}$ National Institute of Science Education and Research, Homi Bhabha National Institute, Jatni, India\\
$^{82}$ National Nuclear Research Center, Baku, Azerbaijan\\
$^{83}$ National Research and Innovation Agency - BRIN, Jakarta, Indonesia\\
$^{84}$ Niels Bohr Institute, University of Copenhagen, Copenhagen, Denmark\\
$^{85}$ Nikhef, National institute for subatomic physics, Amsterdam, Netherlands\\
$^{86}$ Nuclear Physics Group, STFC Daresbury Laboratory, Daresbury, United Kingdom\\
$^{87}$ Nuclear Physics Institute of the Czech Academy of Sciences, Husinec-\v{R}e\v{z}, Czech Republic\\
$^{88}$ Oak Ridge National Laboratory, Oak Ridge, Tennessee, United States\\
$^{89}$ Ohio State University, Columbus, Ohio, United States\\
$^{90}$ Physics department, Faculty of science, University of Zagreb, Zagreb, Croatia\\
$^{91}$ Physics Department, Panjab University, Chandigarh, India\\
$^{92}$ Physics Department, University of Jammu, Jammu, India\\
$^{93}$ Physics Program and International Institute for Sustainability with Knotted Chiral Meta Matter (SKCM2), Hiroshima University, Hiroshima, Japan\\
$^{94}$ Physikalisches Institut, Eberhard-Karls-Universit\"{a}t T\"{u}bingen, T\"{u}bingen, Germany\\
$^{95}$ Physikalisches Institut, Ruprecht-Karls-Universit\"{a}t Heidelberg, Heidelberg, Germany\\
$^{96}$ Physik Department, Technische Universit\"{a}t M\"{u}nchen, Munich, Germany\\
$^{97}$ Politecnico di Bari and Sezione INFN, Bari, Italy\\
$^{98}$ Research Division and ExtreMe Matter Institute EMMI, GSI Helmholtzzentrum f\"ur Schwerionenforschung GmbH, Darmstadt, Germany\\
$^{99}$ Saga University, Saga, Japan\\
$^{100}$ Saha Institute of Nuclear Physics, Homi Bhabha National Institute, Kolkata, India\\
$^{101}$ School of Physics and Astronomy, University of Birmingham, Birmingham, United Kingdom\\
$^{102}$ Secci\'{o}n F\'{\i}sica, Departamento de Ciencias, Pontificia Universidad Cat\'{o}lica del Per\'{u}, Lima, Peru\\
$^{103}$ Stefan Meyer Institut f\"{u}r Subatomare Physik (SMI), Vienna, Austria\\
$^{104}$ SUBATECH, IMT Atlantique, Nantes Universit\'{e}, CNRS-IN2P3, Nantes, France\\
$^{105}$ Sungkyunkwan University, Suwon City, Republic of Korea\\
$^{106}$ Suranaree University of Technology, Nakhon Ratchasima, Thailand\\
$^{107}$ Technical University of Ko\v{s}ice, Ko\v{s}ice, Slovak Republic\\
$^{108}$ The Henryk Niewodniczanski Institute of Nuclear Physics, Polish Academy of Sciences, Cracow, Poland\\
$^{109}$ The University of Texas at Austin, Austin, Texas, United States\\
$^{110}$ Universidad Aut\'{o}noma de Sinaloa, Culiac\'{a}n, Mexico\\
$^{111}$ Universidade de S\~{a}o Paulo (USP), S\~{a}o Paulo, Brazil\\
$^{112}$ Universidade Estadual de Campinas (UNICAMP), Campinas, Brazil\\
$^{113}$ Universidade Federal do ABC, Santo Andre, Brazil\\
$^{114}$ Universitatea Nationala de Stiinta si Tehnologie Politehnica Bucuresti, Bucharest, Romania\\
$^{115}$ University of Cape Town, Cape Town, South Africa\\
$^{116}$ University of Derby, Derby, United Kingdom\\
$^{117}$ University of Houston, Houston, Texas, United States\\
$^{118}$ University of Jyv\"{a}skyl\"{a}, Jyv\"{a}skyl\"{a}, Finland\\
$^{119}$ University of Kansas, Lawrence, Kansas, United States\\
$^{120}$ University of Liverpool, Liverpool, United Kingdom\\
$^{121}$ University of Science and Technology of China, Hefei, China\\
$^{122}$ University of South-Eastern Norway, Kongsberg, Norway\\
$^{123}$ University of Tennessee, Knoxville, Tennessee, United States\\
$^{124}$ University of the Witwatersrand, Johannesburg, South Africa\\
$^{125}$ University of Tokyo, Tokyo, Japan\\
$^{126}$ University of Tsukuba, Tsukuba, Japan\\
$^{127}$ Universit\"{a}t M\"{u}nster, Institut f\"{u}r Kernphysik, M\"{u}nster, Germany\\
$^{128}$ Universit\'{e} Clermont Auvergne, CNRS/IN2P3, LPC, Clermont-Ferrand, France\\
$^{129}$ Universit\'{e} de Lyon, CNRS/IN2P3, Institut de Physique des 2 Infinis de Lyon, Lyon, France\\
$^{130}$ Universit\'{e} de Strasbourg, CNRS, IPHC UMR 7178, F-67000 Strasbourg, France, Strasbourg, France\\
$^{131}$ Universit\'{e} Paris-Saclay, Centre d'Etudes de Saclay (CEA), IRFU, D\'{e}partment de Physique Nucl\'{e}aire (DPhN), Saclay, France\\
$^{132}$ Universit\'{e}  Paris-Saclay, CNRS/IN2P3, IJCLab, Orsay, France\\
$^{133}$ Universit\`{a} degli Studi di Foggia, Foggia, Italy\\
$^{134}$ Universit\`{a} del Piemonte Orientale, Vercelli, Italy\\
$^{135}$ Universit\`{a} di Brescia, Brescia, Italy\\
$^{136}$ Variable Energy Cyclotron Centre, Homi Bhabha National Institute, Kolkata, India\\
$^{137}$ Warsaw University of Technology, Warsaw, Poland\\
$^{138}$ Wayne State University, Detroit, Michigan, United States\\
$^{139}$ Yale University, New Haven, Connecticut, United States\\
$^{140}$ Yonsei University, Seoul, Republic of Korea\\
$^{141}$  Zentrum  f\"{u}r Technologie und Transfer (ZTT), Worms, Germany\\
$^{142}$ Affiliated with an institute covered by a cooperation agreement with CERN\\
$^{143}$ Affiliated with an international laboratory covered by a cooperation agreement with CERN.\\

\end{flushleft} 
  
\end{document}